\documentclass[10pt,a4paper,english]{achemso}
\usepackage[LGR,LGR,LGR,T1]{fontenc}
\usepackage[latin9]{inputenc}
\usepackage{color}
\usepackage{calc}
\usepackage{amsmath}
\usepackage{amssymb}
\usepackage{graphicx}
\usepackage{esint}

\makeatletter

\title{Ultrafast dynamics in the vicinity of quantum light-induced conical
intersections}
\author{András Csehi(1),(2), Markus Kowalewski (3) Gábor J. Halász(4), and
Ágnes Vibók(1),(2)}
\affiliation{(1)Department of Theoretical Physics, University of Debrecen, H-4002
Debrecen, PO Box 400, Hungary}
\affiliation{(2) ELI-ALPS, ELI-HU Non-Profit Ltd, H-6720 Szeged, Dugonics tér
13, Hungary}
\affiliation{(3)Department of Physics, Stockholm University, AlbaNova University
Centre 106 91 Stockholm, Sweden}
\affiliation{(4)Department of Information Technology, University of Debrecen,
H-4002 Debrecen, PO Box 400, Hungary}
\email{vibok@phys.unideb.hu}
\pdfpageheight\paperheight
\pdfpagewidth\paperwidth

\usepackage{babel}

\usepackage{babel}

\usepackage{babel}

\usepackage{babel}

\usepackage{babel}

\makeatother

\usepackage{babel}
\begin{document}
\medskip{}
 \medskip{}
\begin{center}
\fbox{%
\noindent\begin{minipage}[t]{1\columnwidth}%
\begin{center}
{}\includegraphics[scale=0.5]{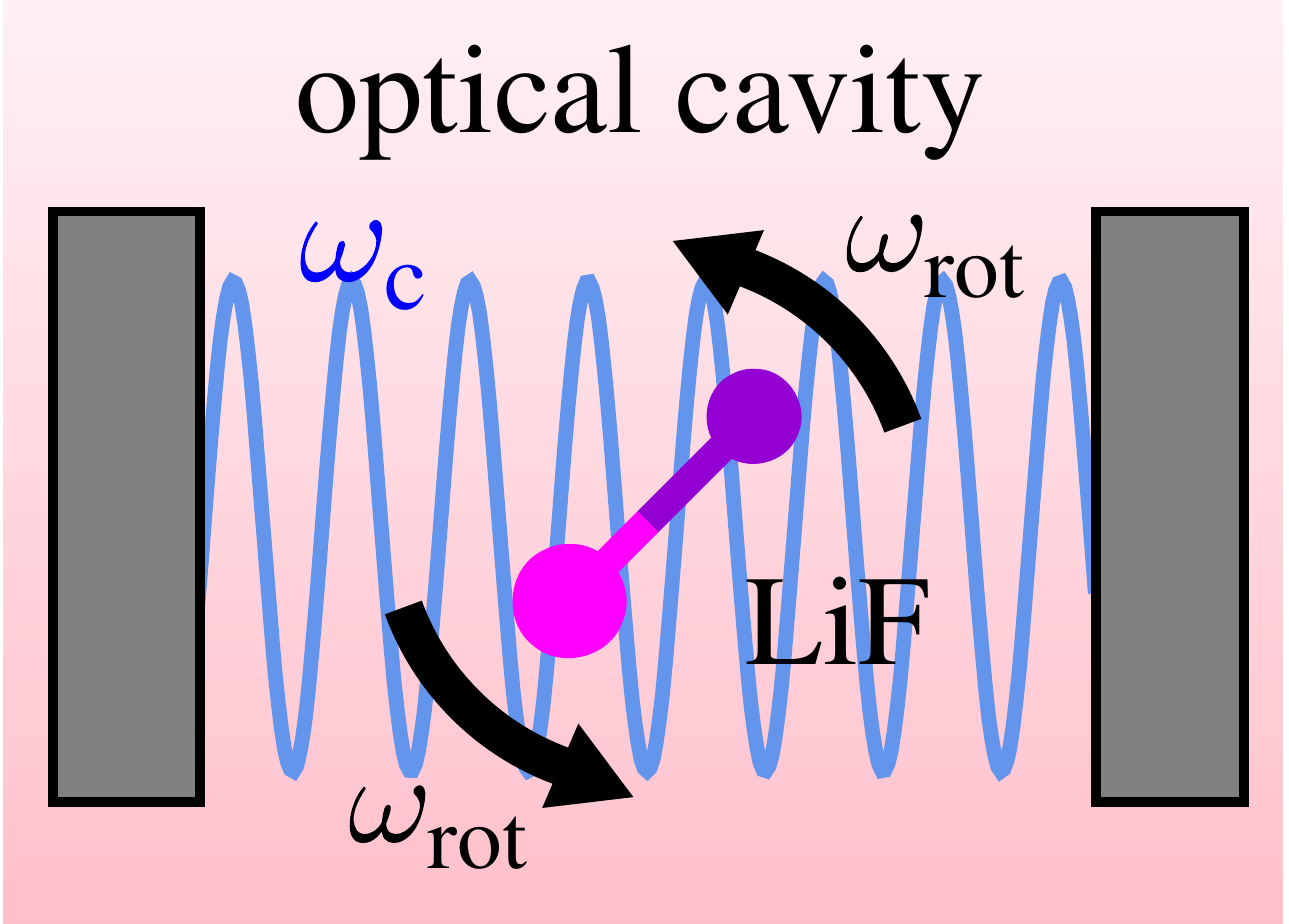} 
\par\end{center}
\begin{center}
{For Table of Contents Only.} 
\par\end{center}%
\end{minipage}} 
\par\end{center}
\begin{abstract}
Nonadiabatic effects appear due to avoided crossings or conical intersections
that are either intrinsic properties in field-free space or induced
by a classical laser field in a molecule. It was demonstrated that
avoided crossings in diatomics can also be created in an optical cavity.
Here, the quantized radiation field mixes the nuclear and electronic
degrees of freedom creating hybrid field-matter states called polaritons.
In the present theoretical study we go further and create conical
intersections in diatomics by means of a radiation field in the framework
of cavity quantum electrodynamics (QED). By treating all degrees of
freedom, that is the rotational, vibrational, electronic and photonic
degrees of freedom on an equal footing we can control the nonadiabatic
quantum light-induced dynamics by means of conical intersections.
First, the pronounced difference between the the quantum light-induced
avoided crossing and the conical intersection with respect to the
nonadiabatic dynamics of the molecule is demonstrated. Second, we
discuss the similarities and differences between the classical and
the quantum field description of the light for the studied scenario. 
\end{abstract}
\begin{description}
\item [{Keywords:}] optical and microwave cavities; Fock states; population
transfer; nonadiabatic couplings; light-induced conical intersections; 
\item [{\textemdash \textemdash \textemdash \textemdash \textemdash \textemdash \textemdash \textemdash \textemdash \textemdash \textemdash \textemdash \textemdash \textemdash \textemdash \textemdash \textemdash \textemdash \textemdash \textemdash \textemdash \textemdash \textemdash \textemdash \textemdash \textemdash \textendash \textemdash \textemdash \textemdash \textemdash \textemdash \textemdash \textemdash \textemdash \textemdash \textemdash \textemdash \textemdash \textemdash \textemdash \textemdash{}}] ~ 
\end{description}
The dynamics initiated in a molecule by absorbing a photon is often
described in the Born-Oppenheimer (BO) or adiabatic approximation
\cite{Born}, where the electronic and nuclear degrees of freedom
are treated separately. However, in some nuclear configurations called
conical intersections (CIs) the mixing between the electronic and
nuclear motions are very significant \cite{Horst1,Baer1,Graham1,Horst2,Baer2}.
Owing to the strong nonadiabatic couplings in the close vicinity of
these CIs the BO approximation breaks down. It is well-known that
these CIs have a significant impact on several important photo-dynamical
processes, such as vision, photosynthesis, molecular electronics,
and the photochemistry of DNA \cite{Domcke,Ashfold1,Kim,Polli1,Todd2}.
During the dynamics the CIs can serve as efficient decay channels
for the ultrafast transfer of the populations. In the following we
call these CIs, which originate from the field free electronic structure,
natural CIs.

Nonadiabatic effects can also appear when molecules are exposed to
resonant laser light. The electric field can couple to two or more
electronic states of the molecule via the non-vanishing transition
dipole moment(s) (TDMs) \cite{Nimrod1,Milan1,Gabor1,Gabor2}. This
results either in a light-induced avoided crossing (LIAC) or a light-induced
conical intersection (LICI) depending on how many nuclear degrees
of freedom are involved in the field induced process \cite{Andris1}.
In case of poly atomic molecules a sufficient number of vibrational
degrees of freedom are always present to span a two-dimensional branching
space (BS), which is indispensable to the formation of LICI. In the
case of diatomics one always has to find a proper second degree of
freedom which can act as a dynamical variable to form a BS. As the
molecule rotates {[}\cite{Tamar1,Herschbach1,Koch1}{]}, the rotational
angle between the molecular axis and the light polarization axis can
serve as the missing degree of freedom for establishing the BS \cite{Yarkony}.
\begin{figure}[p]
\includegraphics[width=0.6\textwidth]{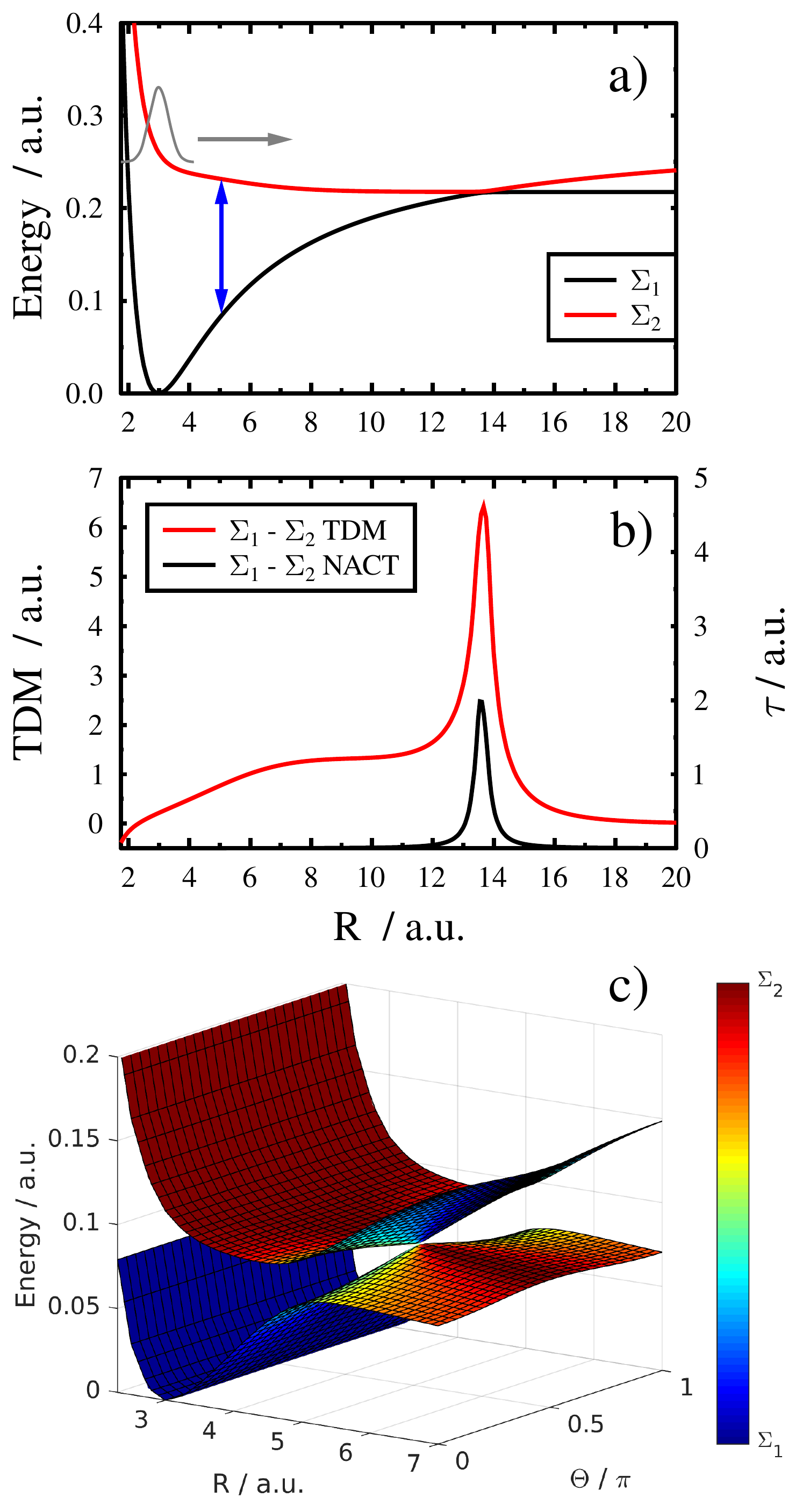}

\caption{\label{fig.1} (a) Bare ground $\Sigma_{1}$ (black line) and excited
$\Sigma_{2}$ (red line) electronic potential energy curves of the
LiF molecule. The initial wave packet around $R=3$\,a.u. is indicated
by the gray line while the resonant coupling of the electronic states
at $R=5$\,a.u. is indicated by the vertical blue arrow. (b) Transition
dipole moment (green line) and intrinsic nonadiabatic coupling (red
line) functions of the $\Sigma_{1}$ and $\Sigma_{2}$ electronic
states. (c) Dressed states potential energy surfaces of the LiF molecule
representing the quantum light-induced CI for a cavity coupling $\chi=0.02$
and a cavity resonance frequency $\omega_{c}=0.1468$\,a.u.. The
color code indicates the state character in the terms of the bare
states $|\Sigma_{1},n+1\rangle$ and $|\Sigma_{2},n\rangle$. The
LICI is located at $\theta$=$\pi$/2. }
\end{figure}

\begin{figure}[p]
\includegraphics[width=0.7\textwidth]{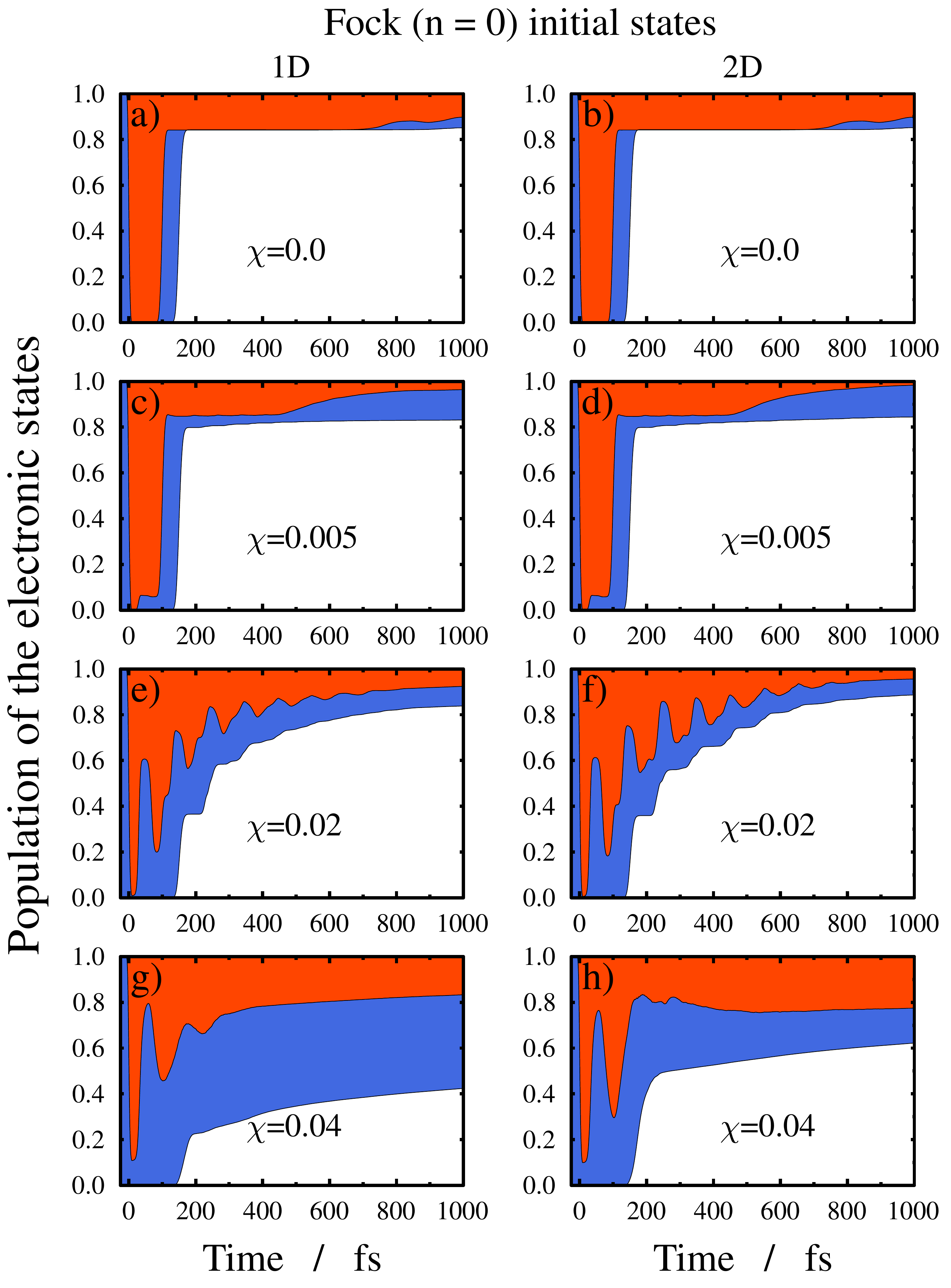}

\caption{\label{fig.2} Population of the ground and excited electronic states
as a function of time computed in 1D and 2D. Panels on the left-hand
side (a)-g)) show the 1D results, while those on the right-hand side
(b)-h)) depict the 2D results for a given cavity coupling strength;
from top to bottom $\chi$=0, $\chi$=0.005, $\chi$=0.02 and $\chi$=0.04.
The blue and red areas correspond to the population of the $\Sigma_{1}$
and the $\Sigma_{2}$ states, respectively. The white regions show
the ground state population being in the dissociation region ($R>20$\,a.u.).
In all the panels the Fock vacuum state ($n=0$) is considered as
an initial state along the cavity mode.}
\end{figure}

\begin{figure}[p]
\includegraphics[width=0.7\textwidth]{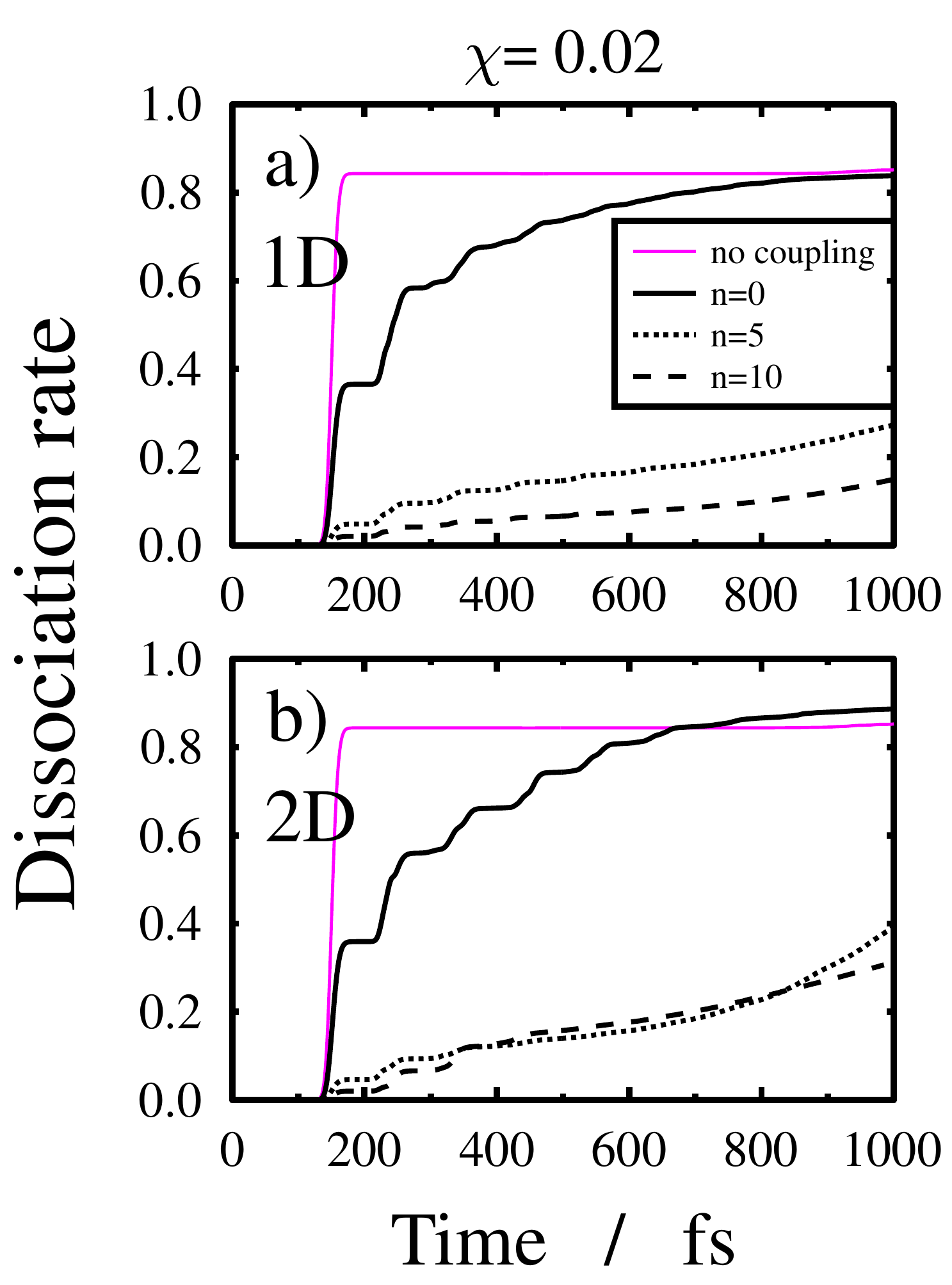}

\caption{\label{fig.3} Time-evolution of the ground state dissociation rate
for the LiF molecule considering different initial Fock states along
the cavity mode. 1D and 2D results are compared in panels a) and b),
respectively for a given cavity coupling strength of $\chi$=0.02.
Besides the different photon number Fock states, the field-free results
are shown by the magenta curves.}
\end{figure}

\begin{figure}[p]
\includegraphics[width=0.7\textwidth]{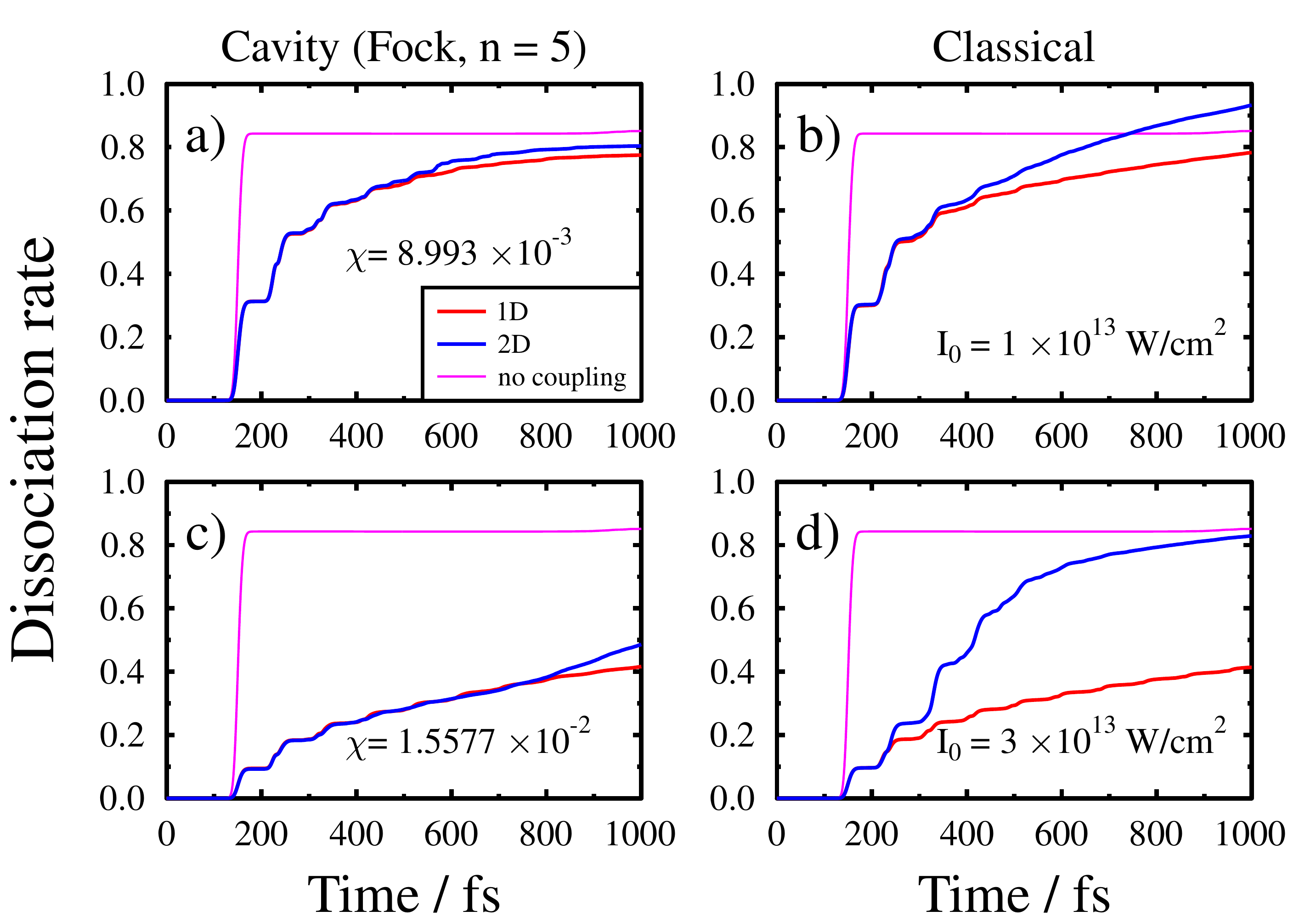}

\caption{\label{fig.4} Comparison of the time-dependent ground-state dissociation
rates of the LiF molecule modified by quantum and classical laser
light. The correspondence between the quantum and classical light
results is demonstrated for two different intensities and cavity coupling
values. The 1D and 2D dissociation rates are shown by the red and
blue lines, respectively. The magenta curves depict the bare results.}
\end{figure}

Nonadiabatic effects can arise in an optical or microwave cavity as
well \cite{Feist1,Feist2,Markus1,Markus2,Markus3}. Describing the
photon-matter interaction with the tool of cavity quantum electrodynamics
(cQED) is an emerging field. It has been successfully demonstrated
both experimentally \cite{Ebbesen1,Ebbesen2,Ebbesen3,Ebbesen4,Ebbesen5,Ebbesen6,Edina}
and theoretically \cite{Feist3,Feist4,Feist5,Spano1,Spano2,Spano3,Yuen1,Yuen2,Yuen3,Rubio1,Rubio2,Rubio3,Rivera1,Palaez1,Triana,Oriol1,Oriol2},
that the quantized photonic mode description of the electromagnetic
field can provide an alternative solution for studying adequately
the light-molecule's quantum control problem. In this framework the
nonadiabatic dynamics arises due to the strong coupling between the
molecular degrees of freedom and the photonic mode of the radiation
field which can alter the molecular levels by controlling the dynamics
of basic photophysical and photochemical processes.

In most of the theoretical descriptions the molecules are treated
via a reduced number of degrees of freedom or by some simplified models
assuming two-level systems. Here, only one vibrational mode which
is strongly coupled to the electronic and photonic degrees of freedom
is taken into account resulting in a new set of ``cavity induced''
or ``polariton'' surfaces in the molecular Hamiltonian \cite{Feist1,Feist2,Markus1,Markus2,Markus3}.
These polariton surfaces can form CIs under special conditions \cite{Markus1}
but are not expected to cross each other in general forming only ``avoided
crossings''. This scenario resembles a one-dimensional (1D), semi-classical
treatment of the light-induced avoided crossing occurring in the nonadiabatic
gas-phase molecular dynamics. Recent studies of the time-dependent
state population in the NaI molecule have successfully demonstrated
that the 1D results obtained by the semi-classical approach \cite{Andris2}
are fully consistent with the results obtained for NaI and quantized
light field \cite{Markus2}.

Adding another molecular degree of freedom, which can either be a
second vibrational mode in case of poly atomics or the rotational
angle between the molecular axis and the polarization axes in the
cavity, one can fully describe the photon-induced quantum dynamics
in the framework of quantum light-induced conical intersection (LICI).
CIs can even be formed in diatomics due to the availability of the
two independent nuclear degrees of freedom, which are essential for
forming a 2D branching space. This picture can now be extended in
a straightforward way from the simplified 1D model with a second molecular
mode to treat the rotational, vibrational, electronic and photonic
degrees of freedom on an equal footing. In the following we call the
latter description the 2D description.

By including the molecular rotation in the cavity treatment, recently
we have investigated the field-dressed rovibronic spectrum of diatomics
in the framework of cavity quantum electrodynamics \cite{Szidarovszky2}.
Incorporating the concept of LICIs with the quantized radiation field,
similar impact on the adiabatic spectrum has been found as in the
case of classical dressed situation \cite{Szidarovszky1}. We have
demonstrated that the ``intensity borrowing'' effect can cause a
significant variation in the pattern of the molecular spectra irrespective
of the origin of the conical intersections. It can be either natural
or light-induced and the latter can be created by classical or quantum
light.

In the present work we go further and investigate theoretically how
the dynamics of the lithium fluoride molecule, which already possesses
a natural avoided crossing, is affected by quantum light-induced CI.
This work complements previous theoretical investigations \cite{Markus2,Andris2},
where only avoided crossings were created by quantum light \cite{Markus2},
but not CIs \cite{Andris2}. Here we investigate the nonadiabatic
quantum dynamics by incorporating the concept of LICIs with the quantized
radiation field.

The aim of this letter is two-fold. First, we would like to study
the quantum light-induced nonadiabatic dynamics of the LiF molecule
both in 1D and 2D frameworks to demonstrate the difference between
the effects of the radiation field-induced avoided crossing (AC) and
CI. The underlying dynamics is mainly governed by the interplay between
one of the quantum light-induced phenomena (either LIAC or LICI) and
of the natural avoided crossing which is present in the field-free
molecule. Second, the similarities and the differences between the
classical and the cavity Fock radiation field description of the light
are investigated. We discuss for both the LIAC and LICI to what extent
the different physical scenarios, a molecule in a laser field and
a molecule in a cavity, show similar or different results.

The general form of the Hamiltonian in the basis of the adiabatic
potential energy surfaces (PES) surfaces $\Sigma_{1}$ and $\Sigma_{2}$
of LiF can be given as 
\begin{equation}
\begin{split}\hat{H}_{cavity} & =\left(-\frac{1}{2M_{r}}\frac{\partial^{2}}{\partial R^{2}}+\frac{1}{2{M_{r}}R^{2}}L_{\theta}^{2}\right){\bf 1}+\begin{pmatrix}V_{\Sigma_{1}} & K\\
-K & V_{\Sigma_{2}}
\end{pmatrix}+\left(-\frac{1}{2}\frac{\partial^{2}}{\partial x^{2}}+\frac{1}{2}\omega_{c}^{2}x^{2}\right){\bf 1}\\
 & +\begin{pmatrix}\chi\omega_{c}\sqrt{2}\cdot\mu_{\Sigma_{1}}\cos{\theta}\cdot x & \chi\omega_{c}\sqrt{2}\cdot\mu_{\Sigma_{1}\Sigma_{2}}\cos{\theta}\cdot x\\
\chi\omega_{c}\sqrt{2}\cdot\mu_{\Sigma_{1}\Sigma_{2}}\cos{\theta}\cdot x & \chi\omega_{c}\sqrt{2}\cdot\mu_{\Sigma_{2}}\cos{\theta}\cdot x
\end{pmatrix}.
\end{split}
\label{eq1}
\end{equation}
Here we assume the electric dipole approximation and that the molecule
interacts with only a single mode of the cavity. In eq.\ \ref{eq1},
the first term represents the rovibrational kinetic energy of the
LiF molecule with R and $\theta$ being the vibrational and rotational
degrees of freedom, respectively. $M_{r}$ is the reduced mass of
the LiF molecule, $L_{\theta}$ is the angular momentum operator (with
$m=0$ fixed) and the ${\bf 1}$ symbol represents the $2\times2$
unit matrix. The second term contains the field-free adiabatic potential
curves $V_{\Sigma_{1}}$ and $V_{\Sigma_{2}}$ (see Fig.\ \ref{fig.1}(a)).
The $K$ field-free nonadiabatic coupling operator operator is approximated
as $K(R)\approx\frac{1}{2M_{r}}(2\tau(R)\frac{\partial}{\partial R}+\frac{\partial}{\partial R}\tau(R))$,
where $\tau(R)$ is first order nonadiabatic coupling term, shown
in Fig.\ \ref{fig.1}(b). The third term in eq.\ \ref{eq1} represents
the harmonic oscillator description of the photon mode with the unit-less
photon displacement coordinate $x$. Here $\omega_{c}$ is the cavity
mode angular frequency. The last term of eq.\ \ref{eq1} describes
the interaction between the molecule and the quantized field. Here
$\chi$ is the cavity coupling strength, while $\mu_{i}$ and $\mu_{ij}$
(i,j=$\Sigma_{1},\Sigma_{2}$) are the permanent and transition dipoles,
respectively. In the actual calculations $\hbar\omega_{c}=3.995$
eV is taken which corresponds to the resonant coupling of the $\Sigma_{1}$
and $\Sigma_{2}$ states around R=5 a.u. The coupling strength $\chi$
ranges from 0.0012 to 0.04 to simulate moderate and strong coupling
strengths. 

The initial wave packet at $t=-10$ fs is created from the product
of the rovibrational ground state of $\Sigma_{1}$ located around
R $\approx$ 3 a.u. and one of the Fock states and is placed on the
lower adiabatic potential $\Sigma_{1}$ . The Fock states are considered
eigenstates of the $\left(-\frac{1}{2}\frac{\partial^{2}}{\partial x^{2}}+\frac{1}{2}\omega_{c}^{2}x^{2}\right)$
Hamiltonian.

The form of the time-dependent Hamiltonian in the classical framework
in the basis of the adiabatic states $\Sigma_{1}$ and $\Sigma_{2}$
of LiF reads: 
\begin{equation}
\begin{split}\hat{H}_{classical}(t) & =\left(-\frac{1}{2M_{r}}\frac{\partial^{2}}{\partial R^{2}}+\frac{1}{2{M_{r}}R^{2}}L_{\theta}^{2}\right){\bf 1}+\begin{pmatrix}V_{\Sigma_{1}} & K\\
-K & V_{\Sigma_{2}}
\end{pmatrix}\\
 & -\varepsilon_{0}\cdot\cos(\omega_{c}t)\cdot f(t)\begin{pmatrix}\mu_{\Sigma_{1}}\cos{\theta} & \mu_{\Sigma_{1}\Sigma_{2}}\cos{\theta}\\
\mu_{\Sigma_{1}\Sigma_{2}}\cos{\theta} & \mu_{\Sigma_{2}}\cos{\theta}
\end{pmatrix}.
\end{split}
\label{eq2}
\end{equation}
In eq.\ \ref{eq2}, the first term represents the vibrational and
rotational kinetic energy (the same as in eq.\ \ref{eq1}), while
the second term contains the field-free $V_{\Sigma_{1}}$ and $V_{\Sigma_{2}}$
potential curves and the $K$ nonadiabatic coupling operator. The
third term of eq.\ \ref{eq2} describes the laser-molecule interaction
in the dipole approximation. Here $\varepsilon_{0}$ is the amplitude
of the electric field, $\omega_{c}$ is the angular frequency of the
laser, $f(t)$ is the envelope function which is set to unity during
the whole propagation (t$_{final}$=1000\,fs). $\mu_{i}$ and $\mu_{ij}$
(i,j=$\Sigma_{1},\Sigma_{2}$) are the permanent and transition dipoles,
respectively. The actual value of the applied laser energy was set
to $\hbar\omega_{c}=3.995$\,eV, which resonantly couples the electronic
states at $R=5$\,a.u., and the peak laser intensity ranges from
$I_{0}=3\times10^{11}$W/cm\,$^{2}$ to $3\times10^{13}$\,W/cm$^{2}$. 

In both the cavity and classical calculations a linearly polarized
resonant pump pulse is applied to initiate the dissociation dynamics.
The form of this pump pulse is given as $\varepsilon_{pump}\cdot\cos(\omega_{pump}t)\cdot g(t)$
where $\varepsilon_{pump}$ is the peak electric field strength, $\omega_{pump}$
is the resonance angular frequency and $g(t)=cos^{2}(\frac{\pi t}{T_{p}})$
(in the {[}-T$_{p}$/2,T$_{p}/2${]} time interval) is the envelope
function. Applying a laser pulse with a center frequency of $\hbar\omega_{pump}=7$
eV, $T_{p}=20$ fs pulse duration, and $I_{pump}=4.8\times10^{13}W/cm^{2}$
peak intensity, 35$\%$ of the total population is excited to $\Sigma_{2}$.
This wave packet then starts to oscillate and gradually dissociate
on $\Sigma_{1}$ via the AC and LICI or LIAC.

The MCTDH (multi configurational time-dependent Hartree) method \cite{MCTDH1,MCTDH2}
has been applied to solve the time-dependent Schrödinger-equation
characterized by either eq.\ \ref{eq1} or eq.\ \ref{eq2}. The
$R$ degree of freedom (DOF) was defined on a sin-DVR (discrete variable
representation) grid ($N_{R}$ basis elements for $R=1.6-60$ a.u.).
The rotational DOF, $\theta$, was described by $N_{\theta}$ Legendre-polynomials,
$P_{l}^{m}(\cos\theta)$ with $m=0$ and $l=0,1,...,N_{\theta}-1$.
The photon displacement coordinate $x$ was described by $N_{x}$
Hermite-polynomials, $H_{n}(x)$ with $n=0,1,...,N_{x}-1$. In the
MCTDH wave function representation, these primitive basis sets ($\xi$)
are then used to construct the single particle functions ($\phi$)
whose time-dependent linear combinations form the total nuclear wave
packet ($\psi$) 
\begin{equation}
\begin{split}\phi_{j_{q}}^{(q)}(q,t)=\sum\limits _{i=1}^{N_{q}}c_{j_{q}i}^{(q)}(t)\xi_{i}^{(q)}(q)\ \ \ q=R,\theta,x\\
\psi(R,\theta,x,t)=\sum\limits _{j_{R}=1}^{n_{R}}\sum\limits _{j_{\theta}=1}^{n_{\theta}}\sum\limits _{j_{x}=1}^{n_{x}}A_{j_{R},j_{\theta},j_{x}}(t)\phi_{j_{R}}^{(R)}(R,t)\phi_{j_{\theta}}^{(\theta)}(\theta,t)\phi_{j_{x}}^{(x)}(x,t)
\end{split}
\label{eq3}
\end{equation}

The actual number of basis functions were $N_{R}=1169$, $N_{\theta}=271$
and $N_{x}=100$ for the vibrational, rotational and photon modes,
respectively. The number of single particle functions for the three
DOFs and on both the $\Sigma_{1}$ and $\Sigma_{2}$ electronic states
were ranging from 10 to 50. The values of $n_{R}=n_{\theta}$ and
$n_{x}$ were chosen depending on the actual value of the $\chi$
cavity coupling strength and $I_{0}$ peak laser intensity. In order
to minimize unwanted reflexions and transmissions caused by the finite
length of the R-grid, complex absorbing potentials (CAP) have been
employed at the last $10\,$a.u. of the grid. The total propagation
time was set to $t_{final}=$1000\,fs and the state populations and
dissociation rates were calculated according to 
\begin{equation}
P_{i}(t)=\langle{\psi_{i}(R,\theta,x,t)|\psi_{i}(R,\theta,x,t}\rangle\quad i=\Sigma_{1},\Sigma_{2}\label{eq4}
\end{equation}
and 
\begin{equation}
P_{diss}(t)=\langle{\psi_{\Sigma_{1}}(t)|\Theta(R-R_{D})|\psi_{\Sigma_{1}}(t)}\rangle+2\int_{0}^{t}dt'\langle{\psi(t')|W|\psi(t')}\rangle.\label{eq5}
\end{equation}
In eq.\ \ref{eq5}, $\Theta$ is the Heaviside step function, $R_{D}$=20\,a.u.is
the starting point of the dissociation region, and $-iW$ is the CAP.
To calculate the potential energy, the dipole moment, and the nonadiabatic
coupling (NAC) curves of the LiF molecule (Fig.\ \ref{fig.1}(b)),
the program packages Molpro \cite{Molpro} was used. These quantities
were calculated at the MRCI/CAS(6/12)/aug-cc-pVQZ level of theory
\cite{Quantumchem-imput}. In particular, $\tau(R)$ has been computed
by finite differences of the MRCI electronic wave functions. The number
of active electrons and molecular orbitals in the individual irreducible
representations of the C$_{2v}$ point group were A$_{1}$ $\rightarrow$
2/5, B$_{1}$ $\rightarrow$ 2/3, B$_{2}$ $\rightarrow$ 2/3, A$_{2}$
$\rightarrow$ 0/1.

We start with the dynamics of the bare molecule (no cavity). Figures
\ref{fig.2}(a) and \ref{fig.2}(b) show the time dependent populations
of both the ground and excited states. The length of the red interval
in vertical direction denotes the amount of the population in the
excited states. The initial set up can be seen at $t<-10$\,fs, when
the whole population is on the ground state. Then a linearly polarized
resonant pump pulse (the details of it are given in the introduction)
is applied to initiate the dissociation dynamics. From now on all
population will be related to the one which is transferred to the
upper states during the pump process. The wave packet of the transferred
population starts to oscillate on the excited state potential curve
and reaches the avoided crossing at t$\approx$80\,fs. Here about
80\% of the population is transferred back to the ground state due
to the intrinsic nonadiabatic coupling. The area colored in blue shows
the amount of the population on the ground state, which begins to
dissociate at t$\approx$140\,fs. At t$\approx$180\,fs approximately
80\% of the population dissociates (marked in white color) and does
not show up again in the dynamics. In the bare molecule the Hamiltonians
for 1D and 2D framework are effectively the same, which results in
identical dynamics. The bumpy structure on the curves at t$\approx$800\,fs
denotes that the wave packet has returned to the avoided crossing
region.

By switching on the cavity coupling between the photonic and the molecular
degrees of freedom, the quantum light-induced nonadiabaticity now
competes with the intrinsic one. In the 1D calculations only the vibrational
molecular degree of freedom mixes with the photonic mode and the electronic
states, creating an avoided crossing between the polariton states.
In the 2D scheme both the rotational and vibrational modes are accounted
for and a LICI can be formed even for a diatomic molecule. We begin
by investigating the vacuum state of the cavity mode ($n=0$). We
performed 1D and 2D calculations for several values of the cavity
coupling and the results of the calculation are shown in Fig.\ \ref{fig.2}.
It can be seen that with increasing coupling strength $\chi$ the
1D and 2D results increasingly deviate from each other and the impact
of the LICI becomes more prominent. For a coupling parameter of $(\chi=0.02)$
large amplitude oscillations can be observed between the electronic
surfaces both in the 1D and 2D populations. This is due to the increased
splitting at the LIAC or LICI for larger coupling strengths creating
decoupled dressed state surfaces \cite{Markus1,Feist1}. However,
at the largest coupling parameter $(\chi=0.04),$ owing to the different
shapes of the upper polaritonic surface in different directions. This
period of this oscillation strongly depends on the molecular orientation
which \textendash{} after a few periods \textendash{} leads to a wash
out of the fingerprints of the oscillation from the total populations
depicted in the figure. The dissociation itself is less suppressed
in the cavity LICI picture than in the case of the quantum light-induced
avoided crossing. This is clear evidence that in the cavity the bond
hardening effect \cite{Bond} is less efficient in 2D than in 1D.
That is, even when the quantized radiation field couples to the molecular
degrees of freedom, the ultrafast decay channel created by the LICI
is more efficient with respect to the fast population transfer than
the avoided crossing. Although the time evolution of the quantum dynamics
is determined by the interplay between the radiation-field induced
and intrinsic nonadiabaticity, at sufficiently strong coupling region
the quantum LICI dominates the process. Similar effects have been
found for natural avoided crossings, natural CIs, and for LIACs and
LICIs induced by classical light field. In this work we have demonstrated
the dynamics of quantum light-induced avoided crossings and quantum
light-induced CIs for the first time.

To better understand the differences between the 1D and 2D results,
we compare the dissociation rates for different photon number of Fock
states ($n=0$, 5, and 10) in Fig.\ \ref{fig.3}. Again, the dissociation
rate in the bare molecule ($\chi=0$) is the same in 1D and 2D. However,
for the vacuum Fock state ($n=0$) a minor difference between the
two schemes can be recognized. The structure of the dissociation curves
is similar but the effect, that the dissociation is more efficient
in 2D, can already be seen here.\textcolor{red}{{} }The efficiency of
LICI is increasing compared to the 1D model when we use Fock states
with $n=5$ and $n=10$ as an initial condition. Increasing the photon
number of the initial state the dissociation rate of the 2D model
less supressed than that of the 1D one. The decay channel provided
by the quantum LICI is more efficient for transferring the population
to the lower polaritonic state, which leads to higher dissociation
rates. The effect is similar to what we experience in the case of
increasing coupling strengths. Namely, the increasing photon number
provides an increasing coupling strength, which is enhanced by $\sqrt{n+1}$,
leading to a more efficient LICI. The rotation of the molecule starts
slowly, therefore up to $t=220$\,fs the 1D and 2D curves are practically
the same for all the studied photon number states. We now discuss
Fig.\ \ref{fig.3}(b) where we see two crossings between the $n=5$
and $n=10$ dissociation curves. Because the bond hardening effect
is stronger for $n=10$, evidently the dissociation starts slower.
At $t=330$\,fs the effect of rotation suddenly becomes apparent
resulting in crossings in the 2D curves (for $n=5$ and $10$). Between
the period of $t=350$\,fs and $t=850$\,fs the rotation already
plays an important role in amplifying the effect of the LICI. This
effect is particularly pronounced for $n=10$.

To gain more insight into this phenomena we have analyzed the two
dimensional wave packet density functions $\left|\psi^{\varSigma_{1}}(R,\theta,t)\right|^{2},$
$\left|\psi^{\varSigma_{2}}(R,\theta,t)\right|^{2},$$\left|\psi^{\varSigma_{1}}(x,\theta,t)\right|^{2}$,
$\left|\psi^{\varSigma_{2}}(x,\theta,t)\right|^{2}$ and found that
the n = 5 and n = 10 Fock states differ significantly from the n=0
vacuum. Namely, a much stronger alignment and bond hardening effect
can be realized in the Fock space including photons than in the case
of a vacuum which then strongly suppresses the dissociation rate.
At weaker coupling strengths the dissociation rates obtained by the
LICI and LIAC are more or less similar, independently from the photon
number of the Fock states, while in the stronger coupling regime the
LICI provides a significantly larger amount of dissociation rate.
In the case of quantum light the molecular degrees of freedom interact
directly with the photonic degrees of freedom and can modify the state
of the light field through stimulated emission. This is not possible
with a classical description of light field.

In Fig.\ \ref{fig.4} we compare the results of the cavity mode for
a photon number $n=5$ with the classical description as obtained
from the 1D and 2D calculations. A similar comparison for the 1D model,
but for a photon number $n=0$, has been made between the quantum
light and classical results of the Na$_{2}$ molecule ground state
population \cite{Andris2}. This is comparable to our earlier findings
where independently from the applied coupling strength practically
no considerable difference has been found between the different 1D
simulations. In contrast, in the 2D calculations significant differences
have been found in the dissociation rates when comparing the classical
and quantum light model. At weaker coupling strengths the difference
is not so prominent, but becomes more significant at stronger couplings.
At really short time dynamics the LIAC and LICI behave similarly but
differ significantly at longer time scale. 

In summary, we could show that the dynamical properties of diatomic
molecules can be strongly modified by quantized light in an optical
cavity. By using the LiF molecule as a showcase example, we demonstrated
that the LICI created by quantum light allows for a more efficient
population transfer than a LIAC given a sufficiently large coupling
strength.. The stronger the cavity coupling, the more prominent the
effect. This difference can be explained by the fact that the LICI
retains a degeneracy between the dressed states even for large coupling
strengths. In contrast, the dressed state curves of the 1D model become
increasingly separated for larger coupling strengths, which leads
to a decreased mixing between the nuclear degrees of freedom and the
electron+photon degrees of freedom. In addition, for the case of LIAC
a close similarity has been found between the classical and the cavity
radiation field description of the light for all the studied coupling
strengths and Fock states. 

A significant difference has been found between the dynamics governed
by the quantum Fock state and the classical light description of the
LICI in the Hamiltonian. The stronger the coupling is, the larger
difference becomes between the dynamics of the LICI formed by quantum
and classical light. By increasing the coupling strength in the cavity
description, the dissociation rate is more and more supressed compared
to the classical 2D calculation. It has also been realized that even
in the quantum description of the LICI the photons which are present
in the Fock state ($n=5$ and $n=10$) can also play an important
role comparing to the Fock vacuum situation. The detailed dissection
of the dynamics and the analysis of the interplay between photonic
degrees and molecular degrees of freedom will be subject of a future
study. 

We note that there is more potential for exploring quantum LICIs in
poly atomic molecules, covering a larger class of chemically relevant
molecules. Here, the cavity induced nonadiabatic molecular dynamics
can be studied in the absence of molecular rotations as they already
provide the necessary number of vibrations to form CIs.

\section*{Acknowledgments}

This research was supported by the EU-funded Hungarian grant EFOP-3.6.2-16-2017-00005.
The authors are grateful to NKFIH for support (Grant K128396). M.K.
acknowledges support from the Swedish Research Council (2018-05346).

\end{document}